\documentclass[11pt]{article}
\usepackage{geometry}                
\usepackage{graphicx}
\usepackage{amssymb}
\usepackage{epstopdf}
\usepackage{amsmath}
\title{Lattice Green functions in all dimensions-version 9/4}
\author{Anthony J Guttmann}

\begin{document}
\maketitle
\abstract{We give a systematic treatment of lattice Green functions (LGF) on the $d$-dimensional diamond, simple cubic, body-centred cubic and face-centred cubic lattices for arbitrary dimensionality $d \ge 2$ for the first three lattices, and for $2 \le d \le 5$ for the hyper-fcc lattice. We show that there is a close connection between the LGF of the $d$-dimensional hypercubic lattice and that of the $(d-1)$-dimensional diamond lattice. We give constant-term formulations of LGFs for all lattices and dimensions. Through a still under-developed connection with Mahler measures, we point out an unexpected connection between the coefficients of the s.c., b.c.c. and diamond LGFs and some Ramanujan-type formulae for $1/\pi.$}

\section{Introduction}

Lattice Green functions arise in numerous problems in condensed matter physics, such as lattice vibration problems, luminescence, diffusion in solids and the dynamics of spin waves \cite{BN70}. They are also central to the theory of random walks on a lattice \cite{H95}, and to the calculation of the effective resistance of resistor networks \cite{C00}.

In the framework of random walks, for a translationally invariant walk on a $d$-dimensional periodic Bravais lattice, a natural question to ask is the probability that a walker starting at the origin of a lattice will be at position ${\vec l}$ after $n$ steps. 
 The probability generating function is known as the Lattice Green Function (LGF). 
 It is
 \begin{equation}
\label{lgf}
P({\vec l};z)=\frac{1}{(2\pi)^d}\int_{-\pi}^{\pi} \cdots  \int_{-\pi}^{\pi} 
 \frac{\exp(-i{\vec l}.{\vec k})d^d{\vec k}}{1-z\lambda({\vec k})} .
 \end{equation}
 The coefficient of $z^n,$ denoted  $[z^n]P({\vec l};z),$ is the probability that a walker starting at the origin will be at position ${\vec l}$ after $n$ steps.
Here $\lambda({\vec k})$ is the {\em structure function} of the lattice, and is given by the discrete Fourier transform of the individual step probabilities. Note that the structure function is not  unique, as a given lattice may usually be represented by more than one set of basis vectors. We give a simple example of this in eqn (\ref{lgf1}) below.

  For example, for the $d$-dimensional hypercubic lattice, the structure function is
 $\lambda({\vec k})=\frac{1}{d}(\cos{k_1}+\cos{k_2} + \cdots \cos{k_d}).$
The probability of return to the origin is $$1 - 1/P({\vec 0};1).$$ In two dimensions we have
     $$P({\vec 0};1)=\frac{1}{(2\pi)^2}\int_{-\pi}^{\pi}  \int_{-\pi}^{\pi} \frac{dk_1\, dk_2}{1-\lambda({\vec k})}. $$
   Since $P({\vec 0};1)$ diverges for two-dimensional lattices, this leads to the well-known result that the probability of return to the origin by a random walker in two dimensions is certain.
   
    Of broader interest are LGFs defined by:
 \begin{equation}\label{lgf0}
P({\vec 0};z)=\frac{1}{(2\pi)^d}\int_{-\pi}^{\pi} \cdots  \int_{-\pi}^{\pi} 
 \frac{d^d{\vec k}}{1-z\lambda({\vec k})} .
 \end{equation}
The coefficient $[z^n]P({\vec 0};z)$ is the probability that a walker starting at the origin returns there after $n$ steps. For a regular lattice of co-ordination number $q,$ an equivalent statement is that $q^n[z^n]P({\vec 0};z)$ is the number of returns to the origin of an $n$-step random walker (and is of course an integer).
For the regular two-dimensional lattices the structure functions are\footnote{Because the honeycomb lattice has two types of site, the expansion parameter $z$ in eqn (\ref{lgf0}) should be replaced by $z^2.$}:
 $$\lambda({\vec k})_{honeycomb}=\frac{1}{9}(1+4\cos^2{k_1}+4\cos{k_1}\cos{k_2}). $$
     $$\lambda({\vec k})_{square}=\frac{1}{2}(\cos{k_1}+\cos{k_2}).  $$
    $$\lambda({\vec k})_{triang}=\frac{1}{3}(\cos{k_1}+\cos{k_2}+\cos{(k_1+k_2)}).  $$
    
    The corresponding LGFs are:
    
    \begin{equation}
    \label{hlgf}
     P({\vec 0};z)_{honey}=\frac{6\sqrt{3}}{\pi(3-z)\sqrt{(3-z)(1+z)}}{\bf K}(k)
    \end{equation}
    where $$k=\frac{4z^2}{(3-z)\sqrt{z(3-z)(1+z)}}$$ and ${\bf K}(k)$ is the complete elliptic integral of the first kind. For readers who prefer hypergeometric representations, recall that $${\bf K}(k) = \frac{\pi}{2} {}_2F_1(\frac{1}{2},\frac{1}{2};1;k).$$
     For the square lattice, the result is remarkably simple,
     \begin{equation}
    \label{sqlgf}
    P({\vec 0};z)_{sq}=\frac{2}{\pi}{\bf K}(z),
    \end{equation}
while for the triangular lattice it is a bit more complicated:
 \begin{equation}
    \label{trlgf}
    P({\vec 0};z)_{tri}=\frac{6}{\pi z\sqrt{c}}{\bf K}(k')
    \end{equation}
where $c=(a+1)(b-1),$ and $$a=\frac{3}{z}+1-\sqrt{3+\frac{6}{z}}, \,\, {\rm and} \,\,  b=\frac{3}{z}+1+\sqrt{3+\frac{6}{z}} $$
 and $$k'=\sqrt{\frac{2(b-a)}{c}}.$$
 
 For the square lattice, we can also use the equivalent structure function 
 $$\lambda({\vec k})_{square}=\cos{k_1}\cos{k_2}. $$ as the square lattice can be considered either as the two dimensional hyper-cubic lattice, which gives the first form of the structure function (above), or as the two-dimensional hyper-body-centred cubic lattice, giving the second form. While the integrands are clearly different, the integrals are equal, demonstrating the point made above that structure functions for a given lattice are not unique. We have
  \begin{equation}\label{lgf1}
4\pi^2P({\vec 0};z)_{square}=\int_{-\pi}^{\pi}   \int_{-\pi}^{\pi} 
 \frac{dk_1 dk_2}{1-\frac{z}{2}(\cos{k_1}+\cos{k_2})} = \int_{-\pi}^{\pi}   \int_{-\pi}^{\pi} 
 \frac{dk_1 dk_2}{1-z(\cos{k_1}\cos{k_2})}.
 \end{equation}
 Similarly, for the honeycomb lattice we can exploit the duality with the triangular lattice and write the structure function as $$\lambda({\vec k})_{honey}=\frac{2}{3}\left (\frac{1}{2}+\lambda({\vec k})_{tri}\right ). $$
 It follows that
   \begin{eqnarray}\label{lgf2}
P({\vec 0};z)_{honey}&=&\frac{1}{(2\pi)^2}\int_{-\pi}^{\pi}   \int_{-\pi}^{\pi} 
 \frac{dk_1 dk_2}{1-\frac{z^2}{3}(1+\frac{2}{3}[\cos{k_1}+\cos{k_2} + \cos(k_1+k_2)])} \\ \nonumber
 & = & \frac{1}{(2\pi)^2}\int_{-\pi}^{\pi}   \int_{-\pi}^{\pi} 
 \frac{dk_1 dk_2}{1-\frac{z^2}{9}(1+4\cos^2{k_1}+4\cos{k_1}\cos{k_2})}. \nonumber
 \end{eqnarray}

 It is also instructive to consider the coefficients in the expansion of the LGFs, as these immediately give the number of returns to the origin after a given number of steps.
 We have:
 \begin{equation}
 P({\vec 0};z)=\sum_{n \ge 0} a_n (\frac{z}{q})^n
 \end{equation}
 where $q$ is the {\it co-ordination number} of the lattice. Thus $q=3$ for the honeycomb lattice, $q=4$ for the square lattice and $q=6$ for the triangular lattice.
 For the honeycomb lattice, 
 \begin{equation}
 a_{2n }=\sum_{j+k+l=n} \left (\frac{n!}{j!k!l!} \right )^2= \sum_{j=0}^n {n \choose j}^2 {2j \choose j}.
 \end{equation}
  For the square lattice,  \begin{equation}
  a_{2n }= {2n \choose n}\sum_{j+k=n} \left (\frac{n!}{j!k!} \right )^2={2n \choose n}^2 . 
  \end{equation}
  For the triangular lattice, 
   \begin{equation} \label{tri}
  a_n=\sum_{j=0}^n {n \choose j}(-3)^{n-j} b_j,\,\,\,  {\rm where} \,\,\, b_{j}=a_{2j}(honeycomb).
  \end{equation}
 
 The result for the triangular lattice appears to be new and has been derived by exploiting the connection between the triangular and honeycomb structure functions, following a similar calculation in \cite{BBBG09} for the diamond-f.c.c. pair.

  \subsection{Watson integrals; $z=1.$}
Setting $z=1,$ $P({\vec 0};1)$ gives the famous {\em Watson integrals} for the $d=3$ case.
     These were first encountered by van Peype\footnote{An even earlier mention is by R Courant in 1928 in the simple-cubic case.}, a student of Kramers, who solved the b.c.c. case, but Watson solved the more difficult s.c and f.c.c cases, and this family of integrals now bears his name.
     The structure functions are:
     $$\lambda({\vec k})_{diam}=\frac{1}{4}(1+(\cos{k_1}\cos{k_2}+\cos{k_2}\cos{k_3} + \cos{k_1}\cos{k_3})).  $$
     $$\lambda({\vec k})_{sc}=\frac{1}{3}(\cos{k_1}+\cos{k_2} + \cos{k_3}).  $$
    $$\lambda({\vec k})_{bcc}=(\cos{k_1}\cos{k_2}\cos{k_3}).  $$
      $$\lambda({\vec k})_{fcc}=\frac{1}{3}(\cos{k_1}\cos{k_2}+\cos{k_2}\cos{k_3} + \cos{k_1}\cos{k_3}).  $$
               
The Watson integrals are given by:
     $$P({\vec 0};1)=\frac{1}{(2\pi)^3}\int_{-\pi}^{\pi}  \int_{-\pi}^{\pi} \int_{-\pi}^{\pi}
 \frac{dk_1\, dk_2\, dk_3}{1-\lambda({\vec k})}, $$
and have been evaluated to yield:
   $$P({\vec 0};1)_{diamond}=\frac{3}{\pi^3}[\Gamma(1/3)]^6=\frac{4}{3}P({\vec 0};1)_{fcc} \approx 1.79288 $$  
     $$P({\vec 0};1)_{sc}=\frac{1}{32\pi^3}(\sqrt{3}-1)[\Gamma(1/24)\Gamma(11/24)]^2 \approx 1.516386 $$ 
     $$P({\vec 0};1)_{bcc}=\frac{1}{2^{14/3}\pi^4}[\Gamma(1/4)]^4 \approx 1.3932039$$ 
 $$P({\vec 0};1)_{fcc}=\frac{9}{4\pi^3}[\Gamma(1/3)]^6 \approx 1.344661 $$  
 
The result for the simple-cubic case is a saga in itself. Watson gave the result in terms of the square of a complete elliptic integral with an algebraic argument in 1939.  Some 40 years later Glasser and Zucker \cite{GZ77} showed that Watson's result could be simplified to
$$P({\vec 0};1)_{sc}=\frac{\sqrt{6}}{96\pi^3}[\Gamma(1/24)\Gamma(5/24)\Gamma(7/24)\Gamma(11/24)]^2,$$ which J Borwein and Zucker simplified to the quoted result some 15 years later in \cite{BZ92}.
 
 \subsection{The general case for arbitrary $z$.}
 More generally, the LGFs for the three-dimensional lattices are also known. For the simple cubic lattice one has:
     $$P({\vec 0};z)=\frac{1}{(\pi)^3}\int_{0}^{\pi}  \int_{0}^{\pi} \int_{0}^{\pi}
 \frac{dk_1\, dk_2\, dk_3}{1-\frac{z}{3}(\cos{k_1}+\cos{k_2} + \cos{k_3})} $$
 Joyce \cite{J98} showed that this could be expressed as
 $$P({\vec 0};z)=\frac{1-9\xi^4}{(1-\xi)^3(1+3\xi)}\left [ \frac{2}{\pi}{\bf K}(k_1)\right ]^2;$$
  where $$k_1^2=\frac{16\xi^3}{(1-\xi)^3(1+3\xi)}; $$  with $$ \xi=(1+\sqrt{1-z^2})^{-1/2}(1-\sqrt{1-z^2/9})^{1/2}.$$

     For the body-centred cubic lattice one has:
     $$P({\vec 0};z)=\frac{1}{(\pi)^3}\int_{0}^{\pi}  \int_{0}^{\pi} \int_{0}^{\pi}
 \frac{dk_1\, dk_2\, dk_3}{1-z(\cos{k_1}\cos{k_2}\cos{k_3})}. $$
 Maradudin {\em et al.} \cite{MMW60} showed that this could be expressed as
 $$P({\vec 0};z)=\left [ \frac{2}{\pi}{\bf K}(k_2)\right ]^2$$
 where $$k_2^2=\frac{1}{2} - \frac{1}{2}\sqrt{1-z^2}.$$

     For the face-centred cubic lattice one has:
     $$P({\vec 0};z)=\frac{1}{(\pi)^3}\int_{0}^{\pi}  \int_{0}^{\pi} \int_{0}^{\pi}
  \frac{dk_1\, dk_2\, dk_3}{1-\frac{z}{3}(c_1c_2+c_1c_3+c_2c_3)} $$
  where $c_i=\cos{k_i}.$
   Joyce \cite{J98}) showed that this could be expressed as
 $$P({\vec 0};z)=\frac{(1+3\xi^2)^2}{(1-\xi)^3(1+3\xi)}\left [ \frac{2}{\pi}{\bf K}(k_3)\right ]^2;$$
 where $$k_3^2=\frac{16\xi^3}{(1-\xi)^3(1+3\xi)};  $$ and $$ \xi=(1+\sqrt{1-z})^{-1}(-1+\sqrt{1+3z}).$$

 Finally, for the diamond lattice one has:
  $$P({\vec 0};z)=\frac{1}{(\pi)^3}\int_{0}^{\pi}  \int_{0}^{\pi} \int_{0}^{\pi}
  \frac{dk_1\, dk_2\, dk_3}{1-\frac{z^2}{4}(1+c_1c_2+c_1c_3+c_2c_3)} ,$$
  which Joyce \cite{J73} pointed out gives
  \begin{equation}
  \label{diamond}
  P({\vec 0};z)=\frac{4}{\pi^2}{\bf K}(k_+){\bf K}(k_-);
  \end{equation}

 where $$k_{\pm}^2=\frac{1}{2} \pm \frac{1}{4}z^2(4-z^2)^{(1/2)} - \frac{1}{4}(2-z^2)(1-z^2)^{(1/2)}.$$
 
 The derivation of these results and some history and erudite discussion can be conveniently found in the book by Hughes \cite{H95}. Note too that as the LGFs can be expressed as the square of a complete elliptic integral, they can also be expressed as the square of a ${}_2F_1$ hypergeometric function. Then by a Clausen type formula they can be expressed as ${}_3F_2$ hypergeometric functions, as shown by Joyce \cite{J94}. We give a slightly simpler representation for the diamond and f.c.c. lattices in Section 3.1
 
 As discussed above for the two dimensional lattices, it is informative to consider the coefficients in the expansion of the LGFs, as these give the number of returns to the origin with a given number of steps.
 We have:
 \begin{equation}
 P({\vec 0};z)=\sum_{n \ge 0} a_n (\frac{z}{q})^n
 \end{equation}
 where $q$ is the {\it co-ordination number} of the lattice. Thus $q=4$ for the diamond lattice, $q=6$ for the simple-cubic lattice, $q=8$ for the body-centred cubic lattice and $q=12$ for the face-centred cubic lattice.
  For the diamond lattice, 
  \begin{equation}
  \label{diam3}
  a_{2n }^{(d)}=\sum_{j+k+l+m=n} \left (\frac{n!}{j!k!l!m!} \right )^2=\sum_{j=0}^n {n \choose j}^2 {2j \choose j}{2n-2j \choose n-j} .
  \end{equation}
  For the simple cubic lattice, 
    \begin{equation}
  \label{sc3}
  a_{2n }^{(sc)}=  \binom{2n}{n}\sum_{j+k+l=n} \left (\frac{n!}{j!k!l!} \right )^2={2n \choose n}\sum_{j=0}^n {n \choose j}^2 {2j \choose j}  .
    \end{equation}
   For the body-centred cubic lattice, 
     \begin{equation}
  \label{bcc3}
   a_{2n }^{(bcc)}= {2n \choose n}^3
     \end{equation}
   while for the face-centred cubic lattice, it was recently pointed out in \cite{BBBG09} that
   \begin{equation} \label{fcc}
   a_{n }^{(fcc)}= \sum_{j=0}^n {n \choose j} (-4)^{n-j} b_j,
   \end{equation}
   where $b_j=a_{2j}^{(d)}$ , given in eqn (\ref{diam3}).

The results presented to date are generally well-known, apart from (\ref{tri}) which is new as far as we know, and the analogous result (\ref{fcc}) which has only recently been obtained. In the next sub-section we discuss the form common to 3-dimensional LGFs, and in the following sections we discuss higher dimensional LGFs which then allows us to consider them in a more general setting, with dimensionality as a parameter.

\subsection{Properties of LGFs for $d < 4.$}
In summary, we see that the two-dimensional LGFs for the standard lattices can be expressed in terms of the complete elliptic integral of the first kind. For three-dimensional lattices, the LGF can be expressed as the product of an algebraic expression and two elliptic integrals of the first kind, with moduli that are algebraic functions of the expansion parameter $z.$  In \cite{G06} Glasser has shown that this is a consequence of the fact that the LGFs for three dimensional lattices can be reduced to the following form:
$$\int_a^b \frac{{\bf K}(u)}{\sqrt{(b-u)(a-u)}}du =\frac{2}{b-a}[\sqrt{(1-a)(1+b)}-\sqrt{(1+a)(1-b)}]{\bf K}(k_+){\bf K}(k_-)$$
where $k_{\pm}$ is an algebraic expression of $a$ and $b.$

Another way of looking at the LGFs of the three-dimensional lattices is in terms of the underlying ODEs. For example, for the s.c. lattice one has \cite{J73}
\begin{equation}
\label{scODE}
4x^2(x-1)(x-9)P^{'''}(x)+12x(2x^2-15x+9)P^{''}(x)+3(9x^2-44x+12)P^{'}(x)+3(x-2)P(x)=0
\end{equation}
where $x=z^2.$ This ODE can be written in the form eqn (\ref{three}) required by Appel's reduction \cite{A80} which gives the solution of certain third-order ODEs in terms of the squares of the solution of an associated second order ODE, notably (\ref{two}). It turns out that {\em all} the 3d LGFs satisfy ODEs with this very special property. The corresponding ODEs for the other lattices are given in Joyce \cite{J01}. In the case of the s.c. lattice, the associated second order ODE is
\begin{equation}
\label{scODE2}
y^{''}(x)+\left [ \frac{1}{x}+\frac{1}{2(x-1)}+\frac{1}{2(x-9)} \right ]y^{'}(x)+\left [ \frac{3(x-4)}{16x(x-1)(x-9)} \right ]y(x)=0.
\end{equation}
In this form, the ODE is readily recognised as the Heun equation, and indeed the solution of (\ref{scODE}) analytic around the origin is
\begin{equation}
\label{scheun}
P_{sc}(z)=[F(9,-3/4;1/4,3/4,1,1/2;z^2)]^2
\end{equation}
where $F$ is a Heun function.

The LGFs of the other three-dimensional lattices (b.c.c, f.c.c, diamond) are also given by third order ODEs which satisfy the Appell condition. The ODEs can be found in Joyce \cite{J01}. For the b.c.c. lattice the reduction to a second order ODE results in the simpler solution in terms of the hypergeometric function,
\begin{equation}
\label{bccheun}
P_{bcc}(z)=[{}_2F_1(1/4,-1/4;1;z^2)]^2.
\end{equation}

For the f.c.c. lattice one again has a solution in terms of the square of a Heun function,
\begin{equation}
\label{fccheun}
P_{fcc}(z)=[F(-3,0;1/2,1,1,1;z)]^2,
\end{equation}
while for the diamond lattice the situation is similar,
\begin{equation}
\label{dheun}
P_{d}(z)=[F(4,-1/2;1/2,1/2,1,1/2;z^2)]^2.
\end{equation}

As an aside we remark that the Heun functions above can all be expressed as a product of an algebraic function and a hypergeometric function, with a complicated algebraic argument. So for example the diamond lattice solution (\ref{dheun}) can be written
$$P_{d}(z)=[\sqrt{4-z^2} - \sqrt{1-z^2}][{}_2F_1(1/2,1/2;1;k_d)]^2$$ where
$$k_d^2 = \frac{1}{2}-\frac{z^2}{4}\sqrt{4-z^2}-\frac{1}{4}(2-z^2)\sqrt{1-z^2},$$ with similar results pertaining for the s.c. and f.c.c. lattices.
This hypergeometric function is immediately recognizable as $\frac{2}{\pi} {\bf K}(k_4).$ 

The value of the LGF at $z=1$ gives the Watson integrals, and they can all be expressed as powers of Gamma functions with rational arguments, times  a numerical factor involving powers of $\pi.$ In two dimensions the analogous integrals are divergent.

The coefficients in the series expansion of the LGF can be expressed as either powers of a binomial coefficient in the simplest case, or as sums of products of binomial coefficients, or as double sums of products of binomial coefficients.

Most attention has been devoted to the calculation of the Watson integrals $P({\vec 0};1),$ but in some circumstances, such as the evaluation of effective resistance in networks of resistors, it is useful to consider the LGF $P({\vec l};1).$ This has been done, by Glasser and Boersma \cite{GB00}, who showed that 
$$P({\vec l};1) = A\cdot P({\vec 0};1) + \frac{B}{\pi^2 P({\vec 0};1) }+ C,$$ where $A, \,\, B, \,\, C$ are lattice-dependent rational numbers, which depend on ${\vec l},$ and satisfy simple recursions. For a discussion of resistor networks using LGFs, see also Cserti in \cite{C00}.

Other generalisations include the anisotropised version of the LGF, in which the structure function for, say, the s.c. lattice is generalised to
$$\lambda({\vec k})=\frac{1}{2+\alpha}(\alpha \cos k_1 + \cos k_2 + \cos k_3).$$ The corresponding LGF is evaluated both at ${\vec 0}$ and at other points in a {\em tour de force} by Delves and Joyce \cite{DJ06}. Other anisotropic structure functions, or linear combinations of structure functions from different lattices, have also been variously treated, see for example Joyce, Delves and Zucker in \cite{JDZ03}. In all cases the results can be expressed as a product of two elliptic integrals. Some insight into why this is so can be found in Glasser's calculation in \cite{G06}, discussed above.
 \subsection{Connections with number theory.}
 In a recent paper, Rogers \cite{R09} obtained new Ramanujan-type formulae for $1/\pi$ by considering the logarithmic Mahler measure of certain 3-variable Laurent polynomials\footnote{In fact the second identity in Rogers's theorem, giving eqn (\ref{pi2}), appeared previously in \cite{C04}.}. Mahler measures are defined and discussed in Section 3.1 below. Theorem 3.1 of Rogers \cite{R09} gives a number of new formulae for $1/\pi$ in terms of sums involving products of binomial coefficients. From equations (\ref{diam3}, \ref{sc3}) above, these can be recognised as coefficients of the LGF of the diamond and simple cubic lattices. Accordingly, we have, as  corollaries, the following surprising formulae for $1/\pi$ in terms of LGF coefficients.
 \begin{equation}
 \frac{2}{\pi} =\sum_{n=0}^\infty (-1)^n \frac{(3n+1)}{32^n} a_{2n }^{(d)}.
 \end{equation}
  \begin{equation}
\label{pi2}
 \frac{8\sqrt{3}}{3\pi} =\sum_{n=0}^\infty  \frac{(5n+1)}{64^n} a_{2n }^{(d)}.
 \end{equation}
  \begin{equation}
 \frac{9+5\sqrt{3}}{\pi} =\sum_{n=0}^\infty (6n+3-\sqrt{3}) \left ( \frac{3\sqrt{3}-5}{4} \right )^n a_{2n }^{(d)}.
 \end{equation}
  \begin{equation}
  \label{rsc}
 \frac{2(64+29\sqrt{3}) }{\pi} =\sum_{n=0}^\infty  (520n+159-48\sqrt{3}) \left ( \frac{80\sqrt{3}-139}{484} \right )^n a_{2n }^{(sc)}.
 \end{equation}
 
One of Ramanujan's 17 formulae for $1/\pi$ \cite{R14} can also be expressed in terms of the number of returns on the b.c.c. lattice.

\begin{equation}
\label{bccr}
  \frac{4}{\pi} =\sum_{n=0}^\infty  \frac{(6n+1)}{256^n} a_{2n }^{(bcc)}.
 \end{equation}
 There is also the remarkable Ramanujan type formula due to the Borwein brothers \cite{BB87} which allows one to calculate any binary digit of $1/\pi$ without calculating the earlier part of the binary expansion. This can be written as:
 \begin{equation}
 \label{bb}
  \frac{16}{\pi} =\sum_{n=0}^\infty  \frac{(42n+5)}{4096^n} a_{2n }^{(bcc)}.
 \end{equation}
 
Note that the general form of a Ramanujan type series for $1/\pi$ is as follows \cite{Z07}:
\begin{equation}
\label{ram}
\alpha f(z_0) + \beta \theta f(z_0) = \frac{1}{\pi}, \,\,\, \theta = z\frac{{\rm d}}{{\rm d}z}
\end{equation}
where $\alpha,$  $\beta$  and $z_0$ are algebraic numbers and $f(z)$ is the analytic solution around the origin of a particularly ``nice" third order Fuchsian linear ODE. As pointed out by Zudilin \cite{Z07}, this ODE must be the (symmetric) square of a second order ODE, a situation first discussed by Appell \cite{A80} more than a century ago, and explained in equations (\ref{three}, \ref{two}) in Appendix A below. As noted in the preceding section, the three dimensional LGFs for all four common 3-d lattices satisfy the Appell property, so in that sense it is not surprising that they are {\em candidates} for a Ramanujan type formula. What is more surprising is that they actually do occur in this way

 For example, the formula given by eqn (\ref{rsc}) satisifes (\ref{ram}) with
 $$\alpha = \frac{1104-591\sqrt{3}}{242}; \,\, \beta=\frac{20(64-29\sqrt{3})}{121}; \,\, f(z_0)=P_{sc}(\vec 0,\frac{3}{11}(5\sqrt{3}-8)).$$
To end this section we point out that there exists another family of formulae for $1/\pi^2,$ called the Ramanaujan-Guillera formulae. Zudilin \cite{Z07} has shown how these may be constructed from Ramanujan type formulae for $1/\pi$ by taking the fourth power of the solution of the associated second-order ODE (\ref{two}) that appears in Appell's reduction. In this way we could construct known formulae for $1/\pi^2$ from the alternative route via the LGF ODEs.

\section{Lattice Green Functions for $d > 3$ and Calabi-Yau differential equations}

Calculating the LGF for $d \ge 4$ is a simple exercise in the case of the hyper-bcc lattice, and rather more difficult in the case of the hyper-f.c.c lattice. In the case of the four-dimensional lattices, a unifying feature is that the LGFs all satisfy a 4th order ODE of Calabi-Yau (C-Y) type. The definition of C-Y ODEs and their properties is given in Appendix B.

Calabi-Yau equations are ODEs that have the formal properties of operators that appear as Picard-Fuchs operators.
More formally, consider a  family of $n$-dimensional CY manifolds\footnote{A Calabi-Yau manifold of dimension $n$ is usually defined as a compact $n$-dimensional K\"ahler manifold satisfying certain technical conditions. A one-dimensional Calabi-Yau manifold is a complex elliptic curve, and is algebraic.}  $X(z)$ parameterized by $z \in C$, where $C$ is an open set on a compact Riemann surface. 
By definition, $X(z)$ has a unique holomorphic $n$-form $\omega(z)$ that vanishes nowhere. The periods $\Omega_i (z) = \int_{\Gamma_i} \omega(z)$ are the integrals of $\omega(z)$ over the $n$-cycles $\Gamma_i$ that form a basis for the $n$-th homology of $X(z)$. As we vary $z$ on $C$, each period $\Omega_i (z)$ will vary in such a way that it satisfies an ODE in $z$. This is the Picard-Fuchs equation for the period $\Omega_i(z)$ of $X(z)$. For further details, see for example \cite{Morrison}.  CY ODEs comprise a class of ODEs that are pivotal in string theory.

   Here we consider only
4$^{\rm th}$ order ODEs, (corresponding to the case of Calabi-Yau threefolds)\footnote{All ODEs considered in this article are linear. To save repetition we delete that adjective, but it should be understood to be implied}. In \cite{AZ07, vEvS07} the authors define C-Y ODE's as (4th order) ODEs satisfying five conditions. The list of known equations is given in \cite{AZES09}, and in an updated form at http://enriques.mathematik.uni-mainz.de/CYequations/

    Consider ODE's of the form
  $$y^{(s)}+a_{s-1}(z)y^{(s-1)}+ \cdots + a_1(z)y'+a_0(z)y(z)=0,$$
  where $\{a_i\}$ are meromorphic fns. of $z.$   If $z=0$ is a regular singular point, we can write $$a_{s-j}(z)=z^{-j}{\tilde a}_{s-j}(z) \,\,\, j=1,\ldots,s,$$ where ${\tilde a}_{s-j}(z)$ are analytic at $z=0.$
   Then the roots of the indicial equation
 $$\lambda(\lambda-1)\cdots (\lambda-s+1)+{\tilde a}_{s-1}(0) \lambda(\lambda-1)\cdots(\lambda-s+2)+$$
 $$ + \cdots + {\tilde a}_{1}(0) \lambda + {\tilde a}_{0}(0)=0$$ determine the exponents of the ODE at the origin.

The first condition is that the ODE must have Maximal Unipotent Monodromy (MUM).
An ODE has MUM if {\em all} the exponents at 0 are zero.

Consider a 4$^{\rm th}$ order ODE which is MUM:
$$y^{(4)}+a_3(z)y^{(3)}+a_2(z)y''+a_1(z)y'+a_0(z)y=0.$$
It has four solutions, $y_0, \,y_1, \, y_2, \, y_3. $

  Being MUM implies that:
\begin{eqnarray}
\label{CY4}
 y_0&=&1+\sum_{n \ge 1} a_n z^n; \\ \nonumber
 y_1&=&y_0\log z+\sum_{n \ge 1} b_n z^n; \\ \nonumber
 y_2&=&\frac{1}{2}y_0\log^2 z+\left (\sum_{n \ge 1} b_n z^n\right ) \log z+ \sum_{n \ge 1} c_n z^n ; \\ \nonumber
y_3&=&\frac{1}{6}y_0\log^3 z+\frac{1}{2}\left (\sum_{n \ge 1} b_n z^n\right ) \log^2 z+ \left (\sum_{n \ge 1} c_n z^n \right ) \log z
 + \sum_{n \ge 1} d_n z^n;
\end{eqnarray}
The second condition is 
\begin{equation}
\label{CY}
a_1=\frac{1}{2}a_2a_3-\frac{1}{8}a_3^3+a_2'-\frac{3}{4}a_3a_3'-\frac{1}{2}a_3^{''}.
\end{equation}

The third condition is that the roots of the indicial equation at $z = \infty,$ $\lambda_1 \le \lambda_2 \le \lambda_3 \le \lambda_4$ are positive rational numbers satisfying $\lambda_1 + \lambda_4 = \lambda_2 + \lambda_3.$ 

The fourth condition is that the power series solution $y_0$ has integral coefficients.

The fifth condition is that the genus zero instanton numbers (defined below) are integral, up to a multiple of a fixed positive integer.

  Define $q = \exp(y_1/y_0) = \sum_{n \ge 1} t_n z^n;$  the inverse function $z=z(q)=\sum u_nq^n$ is the {\em mirror map} in C-Y language. 
 Then the Yukawa coupling $K(q)$ is given by
 $$K(q)=\left ( q \frac{d}{dq} \right )^2\left ( \frac{y_2}{y_0} \right ) = 1+ \sum_{k=1}^\infty \frac{k^3N_kq^k}{1-q^k}.$$
 $N_k$ are called {\em instanton numbers,} and two C-Y equations are considered equivalent if they have the same instanton numbers.
 
 The combinatorial or probabilistic significance of the coefficients of $K(q)$ or of the instanton numbers is not known, except insofar as they are of ``geometric origin", as discussed in Appendix B.

Almkvist et al. \cite{AZES09} have catalogued a large number of 4$^{\rm th}$ order ODEs. 
They, and we, use the operator $$\theta=z\frac{d}{dz},$$ and write the 4$^{\rm th}$ order ODE as 
$${\mathcal D} f(z)=0,$$ where
$${\mathcal D}=\theta^4 +zP_1(\theta) + z^2P_2(\theta) \ldots + z^kP_k(\theta)$$
where $P_l,$ $l=1\ldots k$ are polynomials of degree 4 in $\theta.$

Thus ODEs of 1st degree take the form
$$[\theta^4 +zP_1(\theta)]f(z)=0.$$
They are ODEs of generalised hypergeometric functions.
Almkvist et al  \cite{AZES09} found exactly 14 such ODEs. All can be solved. Their solutions have coefficients expressed as finite sums of products of binomial coefficients. We now consider the C-Y ODEs that arise in the 4-dimensional generalisations of the four common lattices discussed above.

   \subsection{Hyper body-centred cubic lattice}
    
  For the 4-dimensional hyper-bcc lattice, the LGF is
  
   $$P({\vec 0};z)=\frac{1}{(\pi)^4}\int_{0}^{\pi} \cdots \int_{0}^{\pi}
 \frac{dk_1\, dk_2\, dk_3\, dk_4}{1-z(\cos{k_1}\cos{k_2}\cos{k_3}\cos{k_4})}. $$
 
 Expanding $1/(1-\lambda z)$ as a power series in $z,$ which is absolutely convergent for $|z| < \lambda,$ we have
 
$$P({\vec 0};z)=\frac{1}{\pi^4}\sum_{n=0}^\infty z^n \left ( \int_{0}^{\pi}  \cos^n k dk \right )^4$$

$$=\sum_{n=0}^\infty \frac{(\frac{1}{2})_n  (\frac{1}{2})_n(\frac{1}{2})_n(\frac{1}{2})_n}{(1)_n(1)_n(1)_n n!}  z^{2n}$$

$$=\,   {}_4F_3(\frac{1}{2},\frac{1}{2},\frac{1}{2},\frac{1}{2};1,1,1;z^2) = \sum_{n=0}^\infty \binom{2n}{n}^4 \left ( \frac{z}{16}\right )^{2n}$$
 
 This admits to no further simplification, or special values at $z=1,$ though the hypergeometric function is rapidly evaluable, giving 
 $$P({\vec 0};1)\approx 1.1186363871641870683496192575256409167948575515294.$$
 
 The LGF satisfies a first-degree, 4th order  C-Y ODE which is number 3 on the list of Almkvist {\em et al.}

 With $\theta=z\frac{d}{dz},$ the LGF satisfies ${\mathcal D} P({\vec 0};16z)=0,$ where

$${\mathcal D} =\theta^4-256z(\theta+\frac{1}{2})^4.$$

Because of the simple structure, we can also easily treat higher dimensions.
For arbitrary $d$ we have 
$$P({\vec 0};z)=\, {}_dF_{d-1}(\frac{1}{2},\frac{1}{2},\frac{1}{2},\ldots,\frac{1}{2};1,\ldots,1;z^2) $$
$$= \sum_{n=0}^\infty \binom{2n}{n}^d \left ( \frac{z}{2^d}\right )^{2n}$$
which satisfies a $d^{\rm th}$ order Fuchsian ODE,
\begin{equation}
\label{iwan}
\left (\theta^n - z2^n(1 + 2\theta)^n \right ) P_d({\vec 0};2^dz)=0. 
\end{equation}

For $d=5$ this equation is of the form required by the Yifan Yang pullback \cite{A06}, which means that the solutions of the $5^{th}$ order equation can be expressed in terms of the solutions of a $4^{th}$ order equation. This is made more precise in Appendix A. For this particular ODE, given by eqn (\ref{iwan}) above, at order 5 (corresponding to the 5-dimensional hyper bcc lattice), the details of the YY-pullback are precisely worked through in Section 7 of Zudilin's article \cite{Z08}. The pullback results in equation number 204 in \cite{AZES09}. A similar result is true for the 5-dimensional hyper cubic (\ref{5sc}) and hyper diamond (\ref{5d}) lattice LGFs, as discussed below. That is to say, the 5th order ODEs permit a YY-pullback to 4th order ODEs which are in the cited list.

   \subsection{Hyper cubic lattice}
     For the 4-dimensional hyper-sc lattice, the LGF is
           \begin{equation}
\label{4dsclgf}
P({\vec 0};z)=\frac{1}{\pi^4}\int_{0}^{\pi}\cdots\int_{0}^{\pi}
 \frac{dk_1\, dk_2\, dk_3\, dk_4}{1-\frac{z}{4}(c_1+c_2+c_3+c_4)},
\end{equation}
 where $\,\,c_i=\cos{k_i} .$

 Following the treatment in \cite{GG94}, we use the identity $\frac{1}{\lambda}=\int_0^\infty \exp(-\lambda t) dt,$
 to write the LGF as
 
 \begin{eqnarray*}
 P({\vec 0};z)&=&\frac{1}{(\pi)^4}\int_0^\infty \int_{0}^{\pi}\cdots\int_{0}^{\pi} e^{-t}\prod_{j=1}^4 e^{(ztc_j/4)} dtd{\vec k}\\
 &=&\int_0^\infty e^{-t} I_0^4\left ( \frac{zt}{4} \right )=\sum_{n=0}^\infty a_n z^{2n},
\end{eqnarray*}
since $I_0(z)=\frac{1}{\pi}\int_0^{\pi}e^{z\cos \theta} d\theta.$

Clearly the use of this identity reduces the $d$-dimensional hypercubic LGF to a one-dimensional integral, involving the $d^{th}$ power of the Bessel function $I_0(zt/d).$

In this way, Glasser and Guttmann obtained a series expansion which was used to identify the underlying 4th order ODE of degree 2. It turns out to be 
 equation number 16 on the list of Almkvist {et al.}
   With $\theta=z\frac{d}{dz},$ the LGF satisfies ${\mathcal D} P({\vec 0};8z)=0,$ where
     \begin{eqnarray*}
    {\mathcal D}&=&\theta^4-4z(2\theta+1)^2(5\theta^2+5\theta+2)+ \\
    &+&2^8z^2(\theta+1)^2(2\theta+1)(2\theta+3).
    \end{eqnarray*}

 The solution is given in \cite{AZES09} as
 \begin{eqnarray*}
 a_n&=&\binom{2n}{n}\sum_{j+k+l+m=n} \left (\frac{n!}{j!k!l!m!} \right )^2\\
 &=&\binom{2n}{n}\sum_{k=0}^n \binom{n}{k}^2\binom{2k}{k}\binom{2n-2k}{n-k}.
 \end{eqnarray*}



The first equality follows from a result \cite{GM93} of Glasser and Montaldi, who related the hypercubic LGF to the generating function of the squares of multinomial coefficients, which had been previously given by Guttmann and Prellberg \cite{GP93}.
 
      The second equality is in Almkvist \cite{A07}, who gives a further 15 distinct expressions for $a_n,$ all involving single or double sums of products of binomial coefficients.
    
     Almkvist (private communication) also pointed out that the Yukawa coupling in this case is
      $$K(q)=1+4q+164q^2+5800q^3+196772q^4+ \cdots,$$
      and hence the instanton numbers are:
      $$\{3N_k\}=\{12,\,60,\,644,\,9216,\,157536,\,3083604 \cdots \}$$
  From the work of Glasser and Montaldi and Guttmann and Prellberg, we can write for the $d$-dimensional hyper-cubic LGF
\begin{equation}
\label{hsc}
[(2dz)^{2n}]P_d({\vec 0};z)=\binom{2n}{n} \sum_{k_1+k_2+\ldots k_d=n} \left ( \frac{n!}{k_1!k_2! \ldots k_d! }\right )^2
\end{equation}
 The 5 dimensional LGF satisfies a 5th order ODE of degree 3, being equation number 188 on the list \cite{AZES09}, and which can be ``pulled back'' to a degree 12 ODE of 4th order, that is also C-Y. 
 
 For $d < 5$ the multinomial expression can be simplified to products over binomial coefficients. 
Specifically, writing
 \begin{equation}
 P({\vec 0};z)=\sum_{n \ge 0} a_n (\frac{z}{2d})^n,
 \end{equation}
 we have for $d=2,\, 3, \, 4$ respectively

$$ a_l^{(2)}= {2l \choose l} \sum_{j=0}^l {l \choose j}^2 = {2l \choose l}^2 =  {2l \choose l} {}_2F_1(\frac{1}{2},-l,-l;1;1) $$
$$ a_l^{(3)}= {2l \choose l} \sum_{j=0}^l {l \choose j}^2 {2j \choose j} = {2l \choose l} {}_3F_2(\frac{1}{2},-l,-l;1,1;4)$$
$$a_l^{(4)}= {2l \choose l} \sum_{j=0}^l {l \choose j}^2 {2j \choose j}{2l - 2j \choose l -j} = {2l \choose l}^2 {}_4F_3(\frac{1}{2},-l,-l,-l;1,1, \frac{1}{2}-l;1).$$ No such simplification is known for $d \ge 5.$

However in \cite{GP93} I give the recurrence $$S_n^{(d)} = \sum_{m=0}^n {n \choose m}^2 S_m^{(d-1)},$$ where $$S_n^{(d)} =\sum_{k_1+k_2+\ldots k_d=n} \left ( \frac{n!}{k_1!k_2! \ldots k_d! }\right )^2 $$

So for $d=5$ we can write the multinomial as a double sum:
$$S_n^{(5)}=\sum_{k_1=0}^n \sum_{k_2=0}^{n-{k_1} } \left ( {\frac{n!}{k_1!k_2!(n-k_1-k_2)!}} \right )^2
\binom{2k_1}{k_1} \binom{2k_2}{k_2}.$$

Hence 
\begin{eqnarray}
\label{5sc}
a_n^{(5)}&=&\binom{2n}{n} \sum_{k_1 +\ldots k_5=n} \left ( \frac{n!}{k_1!k_2! \ldots k_5! }\right )^2 \\ \nonumber
& =& \binom{2n}{n}\sum_{k_1=0}^n \sum_{k_2=0}^{n-{k_1} } \left ( {\frac{n!}{k_1!k_2!(n-k_1-k_2)!}} \right )^2
\binom{2k_1}{k_1} \binom{2k_2}{k_2}.
\end{eqnarray}

Joyce \cite{G03} has also studied the LGF of the hypercubic lattice, and obtained a rapidly convergent asymptotic expansion for the coefficients of the $d$-dimensional lattice LGF, as well as elucidating the behaviour of the LGF in the vicinity of the dominant branch-point singularities.

Finally we note the identity for the following integral of the $d^{th}$ power of the modified Bessel function $I_0.$
\begin{equation}
\label{besssc}
\int_0^\infty e^{-t} I_0^d\left ( 2zt \right ) dt=\sum_{n=0}^\infty  {2n \choose n} \sum_{k_1 + k_2 +\ldots k_d=n} \left ( \frac{n!}{k_1!k_2! \ldots k_d! }\right )^2  z^{2n}
\end{equation}
This is  just the generating function for returns on the $d$-dimensional hyper-sc lattice.
The corresponding integral with $I_0$ replaced by $J_0$ produces the same result with an extra factor $(-1)^n$ in the sum on the r.h.s.

  \subsection{Hyper diamond lattice}
  There is some confusion in the literature as to the 4d generalisation of the diamond structure\footnote{I too was confused, and my confusion was resolved by David Broadhurst. A correct discussion can be found at www.hermetic.ch/compsci/lattgeom.htm}.   
      
There are two types of site, and the relevant unit vectors are: $$(\pm1,0,0,0),\,(0,1,0,0),\,(0,0,1,0),\,(0,0,0,1)$$ and $$(\pm1,0,0,0),\,(0,-1,0,0),\,(0,0,-1,0),\,(0,0,0,-1)$$ 
Calculating the structure function by taking the Fourier transform of the individual step probabilities gives

  $$\lambda^2= 2\cos(k_2-k_3) + 2\cos(k_2-k_4) + 2\cos(k_3-k_4)+ 4\cos{k_1}(\cos{k_1}+\cos{k_2}+\cos{k_3}+\cos{k_4})+3.$$
To evaluate the LGF,  I did one of the integrals exactly, and generated the series expansion by doing the remaining 3-dimensional integral within Maple. This is far from the most efficient method, but was adequate for our purpose, as the series for the number of $2n$ step returns was immediately recognizable as the coefficients of the square of the 5-dimensional multinomial coefficients. 
\begin{equation*}
 \sum a_{n}z^{2n}=\sum_{i+j+k+l+m=n} \left( \frac{n!}{i!j!k!l!m!}\right )^2 (z/5)^{2n}.
\end{equation*}

As the corresponding result for the 3-dimensional diamond lattice, and its two-dimensional counterpart, the hexagonal lattice, are given by the squares of the four-dimensional and three-dimensional multinomial coefficients respectively, this is hardly surprising.

I gave the ODE for this in a 1993 paper with Thomas Prellberg \cite{GP93}, and it is simply related to the gen. fn. for 5d staircase polygons! It is
${\mathcal D}P({\vec 0},5z) = 0$ where $${\mathcal D}=
\theta^4-z(35\theta^4+70\theta^3+63\theta^2+28\theta+5)+z^2(\theta+1)^2(259\theta^2+518\theta+285)-225z^3(\theta+1)^2(\theta+2)^2,$$ and is number 34 on the list \cite{AZES09} of 4th order, third degree Calabi-Yau equations.

So using the results of the previous subsection, we immediately have 
\begin{equation}
\label{5d}
a_n= \sum_{k_1+k_2+\ldots k_5=n} \left ( \frac{n!}{k_1!k_2! \ldots k_5! }\right )^2 = \sum_{k_1=0}^n \sum_{k_2=0}^{n-{k_1} } \left ( {\frac{n!}{k_1!k_2!(n-k_1-k_2)!}} \right )^2
\binom{2k_1}{k_1} \binom{2k_2}{k_2}.
\end{equation}
Higher dimensional generalisations are obvious. The $d$-dimensional hyper diamond LGF has coefficients given by the sum of the squares of the $(d+1)$-dimensional multinomial coefficients,
\begin{equation}
\label{dhc}
a_n^{(d)}= \sum_{k_1+k_2+\ldots k_{d+1}=n} \left ( \frac{n!}{k_1!k_2! \ldots k_{d+1}! }\right )^2.
\end{equation}
 The 5th order differential equation for the $d=5$ case is given in \cite{GP93}. It is also given as ODE number 130 in the list \cite{AZES09}, where it is shown that it can be ``pulled back" to a 4th order CY ODE of degree 12.

Broadhurst \cite{B09} has also calculated the Yukawa coupling for the LGF ODEs of the diamond lattices in dimensions $3 < d < 10$ and extracted the instanton numbers $n_k(d)$ of the Lambert series. By numerical experimentation he found that $n_k(d)/k^2$ is an integer for $d>3,$ $k>0,$ and gave explicit polynomial representations in $d$ for $n_k(d)$ for $k < 7,$ and verified the conjecture that these are polynomials of degree $k$ up to $d = 9.$

The $d$-dimensional diamond lattice Green function can also be expressed compactly in terms of Bessel function integrals, as shown by Glasser and Montaldi \cite{GM93}. In fact
\begin{equation}
\label{bessd}
\int_0^\infty t I_0^{(d+1)}(zt)K_0(t) {\rm d}t = P_d({\vec 0};(d+1)z)=\sum_{n=0}^\infty  \sum_{k_1+k_2+\ldots k_{d+1}=n} \left ( \frac{n!}{k_1!k_2! \ldots k_{d+1}! }\right )^2 z^{2n}
\end{equation}

\subsection{Connection between the hyper diamond and hyper cubic lattices}

We have seen that the $d$-dimensional hyper-diamond lattice LGF has coefficients (\ref{dhc}) given by the sum of the squares of the $(d+1)$-dimensional multinomial coefficients. The $d$-dimensional hyper cubic lattice has LGF coefficients (\ref{hsc}) given by $\binom{2n}{n}$ times the sum of the squares of the $d$-dimensional multinomial coefficients. Thus there appears to be a simple relationship between the LGFs of the two lattices.

As it happens, this has been formalised in another context by Guttmann and Prellberg \cite{GP93} and Glasser and Montaldi \cite{GM93}. In \cite{GP93} a study was made of the generating functions whose coefficients were the squares of the $d$-dimensional multinomial coefficients, as part of a study of $d$-dimensional staircase polygons. For $d=4$ this generating function was shown to be connected to the LGF of both the (three dimensional) f.c.c. and diamond lattices.

In \cite{GM93} this was clarified, as Glasser and Montaldi showed that the generating function for the $d$-dimensional multinomial coefficients,
\begin{equation}
\label{multinom}
Z_d(x^2)=\sum_{k_1, \ldots , k_d = 0}^\infty \binom{k_1+ \ldots + k_d}{k_1, \ldots ,k_d}^2 x^{2(k_1+ \ldots + k_d)}
\end{equation}
is equivalent to the generating function for the $d$-dimensional hyper cubic LGF

\begin{equation}
\label{hypersc}
P_d(z)=\frac{1}{\pi^d} \int_0^\infty \cdots \int_0^\infty \frac{d\theta_1 \cdots d\theta_d}{1 - \frac{z}{d}[\cos(\theta_1)+ \cdots + \cos(\theta_d)]}
\end{equation}
through an Abel transform. In particular, they showed that
\begin{equation}
\label{forward}
P_d(z)=\frac{2}{\pi}\int_0^1 \frac{Z_d(t^2z^2/d^2)}{\sqrt{1-t^2} }dt
\end{equation}
and
\begin{equation}
\label{backward}
Z_d(x^2)=\frac{d}{dx}\left (x\int_0^1 \frac{tP_d(dxt)}{\sqrt{1-t^2} }dt \right ).
\end{equation}
Here we capitalise on this result by pointing out that $Z_{d+1}(x^2)$ {\em is} the LGF for the $d$-dimensional hyper diamond lattice, with coefficients given by eqn (\ref{dhc}).

Invoking this connection at the coefficient level, implementing eqn (\ref{forward}) only requires the calculation of the integral
$$ \frac{2}{\pi} \int_0^1 \frac{t^{2n}}{\sqrt{1-t^2} } dt = \frac{1}{4^n} \binom{2n}{n}, $$ which reveals the origin of the prefactor $\binom{2n}{n}$
in the expression for the coefficients of the hyper cubic LGF.

Finally, comparison of eqns (\ref{bessd}, \ref{besssc}, \ref{forward}) yields the Bessel function relation
\begin{equation}
\label{connect}
\int_0^\infty e^{-t} I_0^d\left ( zt/d \right ) {\rm d}t = \frac{2}{\pi}\int_0^1 \frac{{\rm d}u}{\sqrt{1-u^2}} \int_0^\infty t I_0^{d}(ztu/d)K_0(t) {\rm d}t.
\end{equation}
Other Bessel function identities discovered through their connection with LGFs can be found in Broadhurst \cite{B08}.

    \subsection{Hyper face-centred cubic lattice}
     
The four-dimensional hyper-fcc lattice Green function is given by

       $$P({\vec 0};z)=\frac{1}{(\pi)^4}\int_{0}^{\pi}\cdots\int_{0}^{\pi}
 \frac{dk_1\, dk_2\, dk_3\, dk_4}{1-\frac{z}{6}\lambda},$$
 where $\lambda= (c_1c_2+c_1c_3+c_1c_4+c_2c_3+c_2c_4+c_3c_4)$ and $c_i=\cos{k_i} .$

In \cite{G09} I found the CY ODE satisfied by $P({\vec 0};24z)$ by doing two of the four integrations exactly, and then expanding the integrand in a power series and integrating term-by-term within Maple. In this way I identified the ODE from a series of some 41 terms, which contains barely any confirmatory coefficients. Subsequently Broadhurst \cite{B09} obtained the series to 100 digits, thus providing ample confirmation of the result.

The two integrals were performed as follows:

Set $a=1-\frac{z}{6}(c_2c_3+c_2c_4+c_3c_4); \;\; b=\frac{z}{6}(c_2+c_3+c_4).$ 

Then the integrand is $[a-b\cos k_1]^{-1}.$

 Use $$\frac{1}{\pi}\int_0^{\pi} \frac{d\theta}{a-b\cos \theta} = \frac{1}{\sqrt{a^2-b^2}}=\frac{1}{\sqrt{(a+b)(a-b)}}$$ to eliminate $k_1.$
 Next write $(a+b)(a-b)=e(c - \cos k_2)(d - \cos k_2),$ where $c, \, d, \, e$ are independent of $k_2,$ and use
 $$\int_0^{\pi}\frac{d\theta}{\sqrt{(c - \cos \theta)(d - \cos \theta)}} = \frac{2 {\bf K}(k)}{\sqrt{(c - 1)(d + 1)}}$$ to eliminate $k_2,$
 where $k^2=\frac{2(c-d)}{(c-1)(d+1)}.$
       We are left with a two-dimensional integral, which was expanded as a power series in $z$ and integrated term-by-term in Maple. We got to 41 terms in a few hours, then searched for an ODE. 
       With $\theta=z\frac{d}{dz},$ the LGF satisfies ${\mathcal D} P({\vec 0};24z)=0,$ where\footnote{This corrects a sign error in the coefficient of $z$ in \cite{G09}. Also, in the list of singularities given there, $-1/18$ should be replaced by $-1/3.$}
\begin{eqnarray*}
{\mathcal D} &=& \theta^4+z(39\theta^4-30\theta^3-19\theta^2-4\theta)\\
&+&2z^2(16\theta^4-1070\theta^3-1057\theta^2-676\theta-192)\\
&-&36z^3(171\theta^3+566\theta^2+600\theta+316)(3\theta+2)\\
&-&2^53^3z^4(+384\theta^4+1542\theta^3+2635\theta^2+2173\theta+702)\\
&-&2^63^3z^5(1393\theta^3+5571\theta^2+8378\theta+4584)(1+\theta)\\
&-&2^{10}3^5z^6(31\theta^2+105\theta+98)(1+\theta)(\theta+2)\\
&-&2^{12}3^7z^7(\theta+1)(\theta+2)^2(\theta+3).
\end{eqnarray*}
This is a 4th order, degree 7 Calabi-Yau ODE with regular singular points at 0, 1/24, -1/3, -1/4, -1/8, -1/12, and $\infty.$
It was new at the time it was found, and is one of only a few known C-Y ODEs of degree 7. It is now ODE number 366 in the list of Almkvist et al. \cite{AZES09}.
 
We do not yet have a nice expression for the series coefficients in terms of binomial coefficients. A result is given in \cite{AZES09}, which we have simplified a little, giving
\begin{eqnarray*}
a_n&=&\sum_{i+j+k+l+m=n} \binom{2i}{ i}\binom{2j}{j}\binom{2k}{k}\binom{l+m}{m}\binom{2(l+m)}{l+m}^2\\ \nonumber
&&\binom{n}{2(l+m)}\binom{n-2l-2m}{n-2i-l-m}\binom{2i-l-m}{i-k-l} 
\end{eqnarray*}

In two dimensions there is a simple connection, at the coefficient level, between the coefficients of the triangular lattice LGF and the honeycomb lattice LGF, given by eqn (\ref{tri}). In three dimensions a similar connection exists between the coefficients of the fcc and diamond lattices, see eqn (\ref{fcc}). Unfortunately there seems no such connection between 4d fcc and 4d diamond lattices.

Subsequently David Broadhurst pointed out to me that a computationaly far more efficient way to obtain the series is to expand the integrand directly. The coefficients of $z^n$ are given by readily evaluatable multinomial coefficients, and the integrals are then just integrals of powers of cosines over a half-period, and we know $$ \frac{1}{\pi} \int_0^\infty \cos^{2n} x \, dx = {2n \choose n}/4^n, $$ so the general term can be written down as a five-fold sum over products of a multinomial coefficient and 6 binomial coefficients. He also showed \cite{B09} that the degree of the ODE could be reduced to 6 by considering the auxiliary function ${\tilde F}_4(z)=F_4[z/(1-18z)]/(1-18z)$ where $F_4(z) $ is the 4-dimensional hyper fcc LGF. The resulting ODE has the C-Y properties. In \cite{B09} Broadhurst also calculated the first 100 instanton numbers. The first few of these are $3, \, -4, \, 64, \, -253, \, 4292, \, -25608.$

Indeed, Broadhurst in a heroic calculation \cite{B09} also found the ODE for the 5-dimensional hyper fcc LGF. Disappointingly, it destroys the regularity seen to date, as it is not a 5th order but a 6th order ODE, of degree 13, and lacks the MUM property. By a similar transformation to that used for the 4-d fcc LGF, the degree can be reduced to 12. Attempts at factorizing the differential operator have been unsuccessful.

\subsection{Lattice Green function type integrals}

There are several other structure functions that are ``obvious" generalisations of the 3d lattice structure functions, and these have been investigated similarly.

The first is:
  $$P({\vec 0};z)=\frac{1}{(\pi)^4}\int_{0}^{\pi}\cdots\int_{0}^{\pi}
 \frac{dk_1\, dk_2\, dk_3\, dk_4}{1-z\lambda}= \sum_{n=0}^\infty a_n z^{2n},$$
 where $\lambda= (c_1c_2c_3c_4+s_1s_2s_3s_4)$ and $c_i=\cos{k_i} ,$ $s_i=\sin{k_i} ,$
  Then I find that $P({\vec 0};z)$ is exactly the same as the LGF of the 4-d hypercubic lattice eqn(\ref{4dsclgf}), with coefficients $a_n$ given by:
\begin{equation}
a_{n}={2n \choose n} \sum_{j=0}^n {n \choose j}^2{2j \choose j} {2n-2j \choose n-j}
\end{equation}

A more interesting structure function is $\lambda= (c_1c_2c_3+c_1c_2c_4+c_1c_3c_4+c_2c_3c_4).$
Again we can efficiently calculate the Green function (I drop the adjective ``Lattice" as there is no obvious lattice driving this structure function) by expanding the integrand and integrating term by term. This yields a four-fold sum of products of a multinomial coefficient and four binomial coefficients. Broadhurst, and independently I, showed that the resulting series satisfies an ODE of order 8 and degree 16. The regular singular point at the origin has roots of the indicial equation equal to 0 (4 times), $\frac{1}{3},$ $\frac{2}{3},$ $\frac{1}{2}, (2 \,\,{\rm times}),$ so is not MUM. As this structure function does not correspond to any known lattice, that absence of the MUM property is of no concern.

\section{Constant term (CT) identities}

An alternative way to view the series expansions for the number of distinct returns to the origin of a random walker after $n$ steps, which is given by the (appropriately scaled) LGF, is to ask for a constant term formulation. The CT is manifestly a more obviously combinatorial approach than the integral representation of the LGF. The CT enumerates the number of distinct $n$-step returns to the origin by directly generating
 all possible sums of lattice basis vectors that sum to zero. It is also immediate that the value is an integer - something that is not so obvious from the integral formulation. 
 
 The idea here is to write down a function which when raised to the $n^{th}$ power has a constant term equal to the number of $n$-step returns. 
 For the $d$-dimensional hyper cubic and hyper b.c.c. lattices, only even values of $n$ are used, as returns to the origin must have an even number of steps. For the triangular and hyper f.c.c lattices, all values of $n$ are used. For lattices with two types of site, such as the honeycomb and diamond lattices, all values of $n$ are used, but the result  is the number of $2n$ step returns to the origin.

 For example, for the square lattice the required expression is $$f(x,y)=(x+\frac{1}{x})(y+\frac{1}{y}).$$ Then CT$f(x,y)^{2n} = {2n \choose n}^2.$ 
 Alternatively, if $$g(x,y)=(x+\frac{1}{x} + y+\frac{1}{y}),$$ then CT$g(x,y)^{2n} = {2n \choose n}^2.$ 
 
 For the triangular lattice the relevant function is $$h(x,y)=(x+\frac{1}{x} + y+\frac{1}{y} + \frac{y}{x} +\frac{x}{y}).$$ For the honeycomb lattice one has $$u(x,y)=(1+x+y)(1+1/x+1/y)$$
 
 For the diamond, s.c, b.c.c, f.c.c. lattices respectively the relevant functions are:
 $${(1/x+x+z(y+1/y))(x+1/x+ (y+1/y)/z)} \,\,\,\, {\rm diamond} $$
 $$(x+\frac{1}{x} + y+\frac{1}{y} + z+\frac{1}{z}) \,\,\,\,{\rm s.c.} $$
 $$(x+\frac{1}{x})(y+\frac{1}{y})(z+\frac{1}{z}) \,\, \,\,{\rm b.c.c.}$$ 
$$ (x+\frac{1}{x})(y+\frac{1}{y}) + (x+\frac{1}{x})(z+\frac{1}{z}) +(z+\frac{1}{z})(y+\frac{1}{y})  \,\,\,\, {\rm f.c..c.}$$
 Some of these results were first given by Domb in \cite{D60}.
 
 For the four dimensional diamond, s.c, b.c.c, f.c.c. lattices respectively the relevant functions are:
 $${(1/x+x+zy+z/y+w/x)(x+1/x+ y/z+1/yz+x/w}) \,\,\,\, {\rm 4d\,\, diamond} $$
 $$(x+\frac{1}{x} + y+\frac{1}{y} + z+\frac{1}{z}+ w+\frac{1}{w}) \,\, \,\,{\rm 4d \,\,s.c.} $$
 $$(x+\frac{1}{x})(y+\frac{1}{y})(z+\frac{1}{z})(w+\frac{1}{w}) \,\,\,\, {\rm 4d \,\,b.c..c.}$$ 
$$ (x+\frac{1}{x})(y+\frac{1}{y}) + (x+\frac{1}{x})(z+\frac{1}{z}) + (z+\frac{1}{z})(y+\frac{1}{y})+(w+\frac{1}{w})(x+\frac{1}{x} + y+\frac{1}{y} + z+\frac{1}{z})   \,\,\,\, {\rm 4d \,\, f.c.c.}$$

 These then give constant term formulae for the C-Y ODEs  numbers 34, 16, 3 and 366 respectively in the list of Almkvist et al. \cite{AZES09}.
 
 \subsection{CT formulations and Mahler measure}
 
 There is a significant, but not yet fully understood connection between the CT formulation of LGFs and the Mahler measure of a polynomial. Papers developing aspects of this connection have been brought to my attention by Almkvist, who mentioned Stienstra's work \cite{S05a, S05b}, Glasser who mentioned Boyd's paper \cite{B98} and Zucker who drew my attention to the recent paper by Rogers \cite{R09}.
 
 If one writes a polynomial as
 $$p(z) = \alpha(z-\alpha_1)(z-\alpha_2)\cdots(z-\alpha_n)$$ then the {\it Mahler measure}  $M(p)$ of the polynomial is given by the product of the absolute value of $\alpha$ and those roots with magnitude at least 1. That is to say,
 $$M(p)= |\alpha|\prod_{|\alpha_i| \ge 1} |\alpha_i|.$$
 
 More generally, for a Laurent polynomial $F(x,y,z)$ one normally defines the {\em logarithmic Mahler measure} ${\rm \bf m}(F),$ which we denote lMm, and the {\em Mahler measure} ${\rm \bf M}(F),$  respectively as
 \begin{equation}
 \label{logm}
 {\rm \bf m}(F)=\frac{1}{(2\pi i)^3} \oint \oint \oint _{|x|=|y|=|z|=1} \log |F(x,y,z)| \frac{{\rm d}x}{x} \frac{{\rm d}y}{y} \frac{{\rm d}z}{z}
 \end{equation}
 and 
  \begin{equation}
 \label{M}
 {\rm \bf M}(F)= \exp({\rm \bf m}(F)),
 \end{equation}
 where the generalisation to more variables is obvious.
 
 It is not our purpose here to summarise the contents of the cited papers, but we mention those of direct relevance to this work. Boyd \cite{B98} considers the lMm of the Laurent polynomial $$(x + \frac{1}{x} + y + \frac{1}{y} + 1)$$ which is, apart from the additive factor 1, the CT kernel for the square lattice LGF given above.  Boyd investigates generalisations of the conjectured result of Deninger \cite{D97} that
 $${\rm \bf m}(x + \frac{1}{x} + y + \frac{1}{y} + 1) = \frac{15}{(2\pi)^2}L(E,2)= L'(E,0),$$ where $L(E,s)$ is the $L$-function of the elliptic curve $E$ of conductor 15 that is the projective closure of  $(x + \frac{1}{x} + y + \frac{1}{y} + 1).$ (In fact Deninger's conjecture included a constant multiple, which Boyd has shown to be precisely 1 to more than 25 digits). Boyd goes on to conjecture a number of other evaluations of the lMm of two-variable Laurent polynomials in terms of  $L'(E,0),$ where $E$ is a given elliptic curve.
 
 Rogers \cite{R09} considers two three-variable Laurent polynomials, and studies
  $$h(u) := {\rm \bf m}(x + \frac{1}{x} + y + \frac{1}{y} + z + \frac{1}{z} +u),$$ and
 $$g(u) := {\rm \bf m}\left (4-u+(y + \frac{1}{y})(z + \frac{1}{z})  + (x + \frac{1}{x}) (y + \frac{1}{y}) + (x + \frac{1}{x}) (z + \frac{1}{z})\right ),$$ 
which are clearly related to the CT kernel of the simple cubic and face-centred cubic lattices respectively. By considering these and other Laurent polynomials, and their lMm, Rogers establishes a number of new identities for the hypergeometric function ${}_5F_4,$ and new formulae for $1/\pi,$ some of which we have mentioned above in Section 1.3.

Other results of Rogers can be transferred to our problem. For example, with a trivial change of variable, Theorem 3.1 of Rogers can be written
\begin{equation}
P({\vec 0},z)_{diamond} = \frac{1}{(1-z^2/4)} {} _3F_2\left (\frac{1}{3},\frac{1}{2},\frac{2}{3};1,1;\frac{27z^4}{64(1-z^2/4)^3} \right ).
\end{equation}
Given the known relationship between the diamond and f.c.c. LGFs, we can also write
 \begin{equation}
P({\vec 0},z)_{fcc} = {} _3F_2\left (\frac{1}{3},\frac{1}{2},\frac{2}{3};1,1;\frac{z^2(3+z)}{4} \right ).
\end{equation}

Alternative expressions in terms of the squares of a ${}_2F_1$ hypergeometric function were first given by Joyce in 1994 \cite{J94}, who also gave expressions in terms of different ${}_3F_2$ hypergeometric functions. This would give a slightly different version of Theorem 3.1 of Rogers. The expressions above for the LGFs on the f.c.c. and diamond lattices are slightly simpler than those in \cite{J94}.

Finally, Stienstra's two rich papers \cite{S05a, S05b} make connections between Mahler measures, the partition function of some dimer models, and instanton numbers in string theory. It is known that the square lattice dimer partition function can be related to a certain class of spanning tree generating function, which can in turn be obtained by integration of the LGF \cite{G10}. We will expand on these connections in a future publication.

 \section{Conclusion}
 We  have summarised known results for LGFs in lower dimension, and present less-well-known results for higher dimensional results. These are then combined to give a cohesive view for some lattices as a function of dimensionality (for the $d$-dimensional diamond, simple-cubic and body-centred cubic lattices). For the face-centred cubic lattice, the situation is less satisfactory, though results for $2 \le d \le 5$ are given, but we cannot give an expression that holds for arbitrary dimensionality, except through the constant-term formulation.
 
 For two dimensional lattices, the LGF can be expressed in terms of the complete elliptic integral ${\bf K}(k).$ 
 For three dimensional lattices, the LGF can be expressed in terms of products of two complete elliptic integral ${\bf K}(k_-){\bf K}(k_+),$ or as the square of a ${}_2F_1$ hypergeometric function or as the first power of a ${}_3F_2$ hypergeometric function. We find slightly simpler ${}_3F_2$ representations for the diamond and f.c.c. lattices.

Joyce \cite{J94} also found a remarkable result in that it is possible to express the LGF for the four common three-dimensional lattices discussed above in terms of the LGF for the two-dimensional honeycomb lattice, (\ref{lgf2}). First, define $$R(\xi) = P_{honey}(\vec 0,3z).$$ The series expansion of $R(\xi)$ has integral coefficients giving the number of $2n$ step returns to the origin of a random walker on the honeycomb lattice. Then $$ P_{fcc}(\vec 0,z) \equiv  \hat P_{fcc}(\vec 0,\xi) = (1-3\xi^2)^2[R(\xi) ]^2$$ where $$z \equiv z(\xi) = -12\xi^2/(1-3\xi^2)^2.$$ For the simple-cubic lattice Joyce finds $$ P_{sc}(\vec 0,z) \equiv  \hat P_{sc}(\vec 0,\xi) = (1-9\xi^4)[R(\xi) ]^2$$ where $$z^2 \equiv z^2(\xi) = 36\xi^2(1 - 9\xi^2)(1- \xi^2)/(1-9\xi^4)^2.$$ For the b.c.c. lattice he finds $$ P_{bcc}(\vec 0,z) \equiv  \hat P_{bcc}(\vec 0,\xi) = (1-9\xi^2)^{1/2}(1-\xi^2)^{3/2}[R(\xi) ]^2$$ where $$z^2 \equiv z^2(\xi) = -64\xi^6/[(1 - 9\xi^2)(1- \xi^2)^3].$$ Finally for the diamond lattice Joyce gives $$ P_{d}(\vec 0,z) \equiv  \hat P_{d}(\vec 0,\xi) = (1-9\xi^2)(1-\xi^2)[R(\xi) ]^2$$ where $$z^2 \equiv z^2(\xi) = -16\xi^2/[(1 - 9\xi^2)(1- \xi^2)].$$

 For four dimensional lattices, the LGF can be expressed as an integral with an integrand that is a product of an algebraic function and a product of two complete elliptic integrals ${\bf K}(k)$ for  the s.c.  lattice by combining eqns (\ref{diamond}) and (\ref{forward}), giving 
 $$ P_{4d\,\,s.c.}= \int_0^1 \frac{{\bf K}(k_+){\bf K}(k_-){\rm d}x}{\sqrt{1-x^2}}; $$  a similar representation can be found for the 4d b.c.c. lattice. It remains to be seen whether this can be achieved for the 4d f.c.c. and diamond lattices. More generally, it still remains to be seen if the results for the 4d LGFs can be presented in a unified manner, as Joyce  has done for the 3d lattices, as described in the preceding paragraph. What is clear however is that the remarkable results that pertain for the 3-d lattices are due to the fact that the underlying 3rd order ODE has the almost-magical Appell reduction property, allowing the solutions to be expressed in terms of the solutions of an associated second-order ODE. For the 4-d LGFs, we have the almost equally remarkable property that the underlying ODEs are all of Calabi-Yau type. For the 5-d LGFs, this beautiful property appears to be broken by the 5-d f.c.c lattice, though it holds for the other three lattices, all of which have solutions that can be expressed in terms of an associated fourth order ODE.
 
 For the $d$-dimensional s.c.  lattice, we can write the result for arbitrary dimensionality in terms of an integral over the $d^{th}$ power of a modified Bessel function, given by (\ref{besssc}), with a corresponding result for the $d$-dimensional diamond lattice, given by (\ref{bessd}). The corresponding result for the $d$-dimensional b.c.c.  lattice is even simpler, being expressible in terms of a ${}_{d}F_{d-1}$ hypergeometric function, or more simply still we can write the number of $2n$ step returns as ${2n \choose n}^d.$ We have no corresponding result for the hyper-fcc lattice.
 
 For the two-dimensional triangular lattice, the coefficients can be expressed in terms of those of the honeycomb lattice, as shown in eqn (\ref{tri}). For the three-dimensional f.c.c. lattice, the coefficients can be expressed in terms of those of the diamond lattice, as shown in eqn (\ref{fcc}). Unfortunately, it seems that for the 4d f.c.c. lattice, there is no simple expression in terms of the coefficients of the 4d diamond lattice. More precisely, we have been unable to find one.
 
 The number of returns has been given by a constant-term formula for all lattices, incidentally augmenting the table in \cite{AZES09}, and it is only this formulation that allows us to generalise the f.c.c. lattice to arbitrary dimension. That is to say, we can write down an expression for the number of returns of arbitrary length on any regular lattice of $d$-dimensions. Unfortunately, this still leaves one with a non-trivial computational problem in seeking the numerical values.
 
 We also find that some remarkable Ramanujan-type formulae for $1/\pi$ involve the coefficients of the 3-dimensional diamond, s.c and b.c.c LGFs.

\section*{Acknowledgements}
I would like to thank Geoff Joyce for introducing me to Lattice Green Functions exactly 40 years ago when we shared an office, and Larry Glasser with whom I have corresponded and occasionally collaborated for many years. More recently I have benefited from discussions with Gert Almkvist, Richard Brak, David Broadhurst, David Bailey, Jon Borwein, Omar Foda, Ole Warnaar, John Zucker and Wadim Zudilin, and I would like to thank them all. Iwan Jensen and Gary Iliev kindly helped me manipulate some ODEs with Maple. Significant corrections to an earlier draft of the manuscript were made by Gert Almkvist,  Omar Foda, Larry Glasser, Bernie Nickel and Wadim Zudilin, for which I am grateful. The information in footnotes 3 and 8 was kindly supplied by Mathew rogers and Gert Almkvist respectively. This work is supported by the ARC through a Discovery Grant, and it is a pleasure to acknowledge their support.

\section*{Appendix A--Yifan Yang pullback}
In this appendix we discuss the notion of YY-pullback (a contraction of the eponymous Yifan Yang pullback). In preparing this appendix we have drawn heavily on the papers by Almkvist \cite{A06} and by Almkvist, Van Straten and Zudilin \cite{ASZ09}. First, recall the result of Appell, \cite{A80}, established 130 years ago, that if a 3rd order ODE can be written as
\begin{equation}
\label{three}
f'''(x)+3P(x)f''(x)+[2P^2(x)+P'(x)+4Q(x)]f'(x)+[4P(x)Q(x)+2Q'(x)]f(x)=0,
\end{equation}
 then its solution can be expressed in terms of the two linearly independent solutions, $g_1(x)$ and $g_2(x)$ of the 2nd order ODE
\begin{equation}
\label{two}
g''(x)+P(x)g'(x)+Q(x)g(x)=0.
\end{equation}
In fact $$f(x)=Ag_1^2(x)+Bg_1(x)g_2(x)+Cg_2^2(x).$$
Equation (\ref{three}) is called the {\em symmetric square} of eqn (\ref{two}). An example of the utility of this connection can be found in \cite{GP93}, while another example establishes Clausen's formula relating the square of a certain $ {}_2F_1$ hypergeometric function to a ${}_3F_{2}$ hypergeometric function.

Consider now a 4th order ODE satisfying the C-Y conditions described by eqn (\ref{CY}) and the four other conditions given in Section 2.
\begin{equation}
\label{fourth}
h^{(iv)}(x)+P(x)h'''(x)+Q(x)h''(x)+R(x)h'(x)+S(x)h(x) = 0.
\end{equation}
The CY condition implies that the four linearly independent solutions $h_0, \,\, h_1, \,\,h_2, \,\,h_3$ of (\ref{fourth}) are as in (\ref{CY4}). Then the six Wronskian functions
\begin{equation}
\label{wronsk}
w_{jk} = W(h_j,h_k) = {\rm det} \begin{pmatrix}
h_j & \phantom{-}h_k \\
h'_j & \phantom{-}h'_k \\
\end{pmatrix}
, \,\, 0 \le j < k \le 3,
\end{equation}
are not linearly independent over $\mathbb C.$ Indeed, $$w_{03} = w_{12},$$ a condition equivalent to eqn (\ref{CY}). The Wronskians $xw_{ij}$ satisfy a 5th order ODE: 
\begin{equation}
\label{fifth}
u^{(v)}(x)+\hat P(x)u^{(iv)}(x)+\hat Q(x)u'''(x)+\hat R(x)u''(x)+\hat S(x)u'(x)+\hat T(x)u(x) = 0.
\end{equation}
Let ${\hat w}_0 = xw_{01}$ and ${\hat w}_1 = xw_{02}.$ Then if we construct the Wronskian of these Wronskians, we have $$W({\hat w}_0,{\hat w}_1) = x^2y_0^2 \exp(-\frac{1}{2}\int P(x) {\rm d}x).$$
This allows one to recover the solution $h_0$ from the solutions ${\hat w}_0,\, {\hat w}_1$ of the fifth order equation by using $$P(x) = \frac{2}{x} + \frac{2}{5}\hat P (x).$$ Then multiplying by a suitable factor we get the Yifan-Yang pullback:
$$\tilde y(x) = x^{5/2}\sqrt{W({\hat w}_0,{\hat w}_1) }\exp(-\frac{1}{5}\int \hat P(x) {\rm d}x).$$

 The motivation for this is given in \cite{A06}, but the point is that the YY-pullback is usually of lower degree (typically half the degree) of the ordinary pullback found in the table \cite{AZES09}.

A number of examples are given in \cite{AZES09} and a few are cited in the text above. For example, the 5th order ODE satisfied by the 5-dimensional hypercubic LGF is of degree 3, as shown in entry number 188 of \cite{AZES09} but its fourth order YY-pullback is of degree 6, though is not yet in the Table 1 \cite{A10}.

\section*{Appendix B--Calabi-Yau differential equations}
In this appendix we give an attenuated account of the much richer and more detailed material by Almkvist, van Enckevort, van Straten and Zudilin in \cite{AZ07, AZES09} and \cite{ASZ09}. The name {\it Calabi-Yau differential equations} comes from a search by the just-cited authors for ``arithmetically nice'' differential operators of order 4 and 5. Before discussing what is meant by the adjective, first, following  \cite{ASZ09}, consider the second order, second degree linear ODE,
\begin{equation}
\label{second}
(\theta^2 - z(a\theta^2 + a\theta +b) + cz^2(\theta + 1)^2)f(z) = 0
\end{equation}
where, as usual, $\theta = z\frac{{\rm d}}{{\rm d}z}.$
We require it to have a unique analytic solution $f_0(z) = 1 + \sum a_n z^n$ where the coefficients $a_n$ are integers. There will also be a second solution $f_1(z) = f_0(z)\log z + g(z),$ where $g(z)$ is an analytic function and $g(0)=0.$ Furthermore, we require the function $$q(z) = z\exp(g(z)/f_0(z)) = \sum_{n \ge 1} b_nz^n$$ to also have coefficients $b_n$ that are integers.
In Zagier \cite{Z09}, the result of an exhaustive search for values of $\{a,b,c\}$ satisfying these conditions is reported. A necessary condition is that the values $\{a,b,c\}$ are integers. However there are only 14 non-degenerate sets of triples found.

Moving to third order ODEs, if we consider the third order, second degree linear ODE,
\begin{equation}
\label{third}
(\theta^3 - z(2\theta+1)(\hat a\theta^2 + \hat a\theta +\hat b) + \hat cz^2(\theta + 1)^3)f(z) = 0,
\end{equation}
and if we impose exactly the same conditions as for the second-order equation, it was found by Almkvist et al. \cite{AZES09} that there are again only 14 triples $\{\hat a,\hat b,\hat c\}$ satisfying these conditions. As an aside, it is pointed out in \cite{AZ07} that the second-order equation with the triple $\{11,3,-1\}$ arises in Ap\'ery's proof of the irrationality of $\zeta(2),$ while  the third-order equation with the triple $\{17,5,1\}$ arises in Ap\'ery's proof of the irrationality of $\zeta(3).$ Furthermore Almkvist et al in \cite{ASZ09} prove a natural bijection between the set of triples $\{a,b,c\}$ and the set of triples $\{\hat a,\hat b,\hat c\},$ and use that observation to link the solutions by a rational map.

In order to generalise these results to higher order ODEs (specifically orders 4 and 5), the approach taken in \cite{AZ07, AZES09} was to require the differential operator to satisfy similar conditions to those satisfied by the second- and third-order ODEs just considered. These are:\\
(i) the ODE must be Fuchsian (all singular points are regular),\\
(ii) it has the MUM property (all exponents are zero for the singular point at the origin),\\
(iii) the unique analytic solution at the origin, $f_0(z) = 1 + \sum a_n z^n$ has integral coefficients $a_n,$ \\
(iv) the solution $f_1(z) = f_0(z)\log z + g(z),$ $g(0)=0,$ gives  an expansion with integral coefficients $b_n$ for the function $$q(z) = z\exp(g(z)/f_0(z)) = \sum_{n \ge 1} b_nz^n.$$

The Calabi-Yau condition (\ref{CY}) was additionally imposed in \cite{AZ07, AZES09}, though in hindsight it appears, at least experimentally, to be a consequence of the four conditions above. Likewise the integrality of the (possibly scaled) instanton numbers\footnote{We have been informed \cite{A10} that in as yet unpublished work, Michael Bogner and Stefan Reiter of Mainz have found a counter-example. That is, a fourth-order ODE in which $y_0$ $q$ and $K(q)$ all have integer coefficients, but the instanton numbers have denominators of unlimited size}. By a variety of ingenious constructions and exhaustive searches, there are now more than 350 known equations of degree 4 catalogued in \cite{ AZES09}. They are called {\it Calabi--Yau} since they can often be identified with Picard--Fuchs differential equations for the periods of 1-parameter families of Calabi--Yau manifolds. It is believed that all these ODEs are of geometric origin. Technically, and here we quote \cite{ASZ09}, this means that they correspond (as subquotients of the local systems) to factors of Picard--Fuchs equations satisfied by period integrals for some family of varieties over the projective line.

Taking this connection with geometry a little further, Almkvist et al. show that, of the 14 solutions  in the case of second order ODEs, four are hypergeometric, four are Legendrian and six are sporadic. The hypergeometric solutions correspond to the value $c = 0$ in (\ref{second}). In the Legendrian case one has $c = a^2/4$ in (\ref{second}). In both these cases the differential operators are Picard--Fuchs operators of the extremal rational elliptic surfaces with three singular fibres \cite{MP86}. Finally, the six sporadic cases are also of geometric origin, arising as Picard-Fuchs equations of the six families of elliptic curves with four reduced singular fibres \cite{B82, MP86}.

\end{document}